\documentclass[aps,prmaterials,reprint,superscriptaddress,floatfix]{revtex4-2}

\usepackage{amsmath}
\allowdisplaybreaks[1] 
\usepackage{graphicx}
\usepackage{amssymb}
\usepackage{booktabs}
\usepackage[percent]{overpic}

\graphicspath{{figs/}}

\begin{document}
	
	\title{Electromechanical properties of the $\mathbf{180}^{\boldsymbol{\circ}}$ domain wall in PbTiO$_3$.}
	
	\author{I.~Rychetsky}
	\email{rychet@fzu.cz}
	\author{A.~Klic}
	\affiliation{Institute of Physics of the Czech Academy of Sciences (FZU), Na Slovance 2, 18200 Prague 8, Czech Republic.}
	\author{W.~Schranz}
	\affiliation{University of Vienna, Faculty of Physics, Boltzmanngasse 5, 1090 Wien, Austria.}
	
	\date{\today}
	
	\begin{abstract}
		We analyze the electromechanical response of the $180^{\circ}$ ferroelectric domain wall in tetragonal PbTiO$_3$ by combining first-principles calculations with a Landau--Ginzburg--Devonshire (LGD) description. Using regular multidomain structures with varying domain-wall density, we extract polarization profiles and lattice distortions and map them onto the continuum model to determine conventional (homogeneous) and gradient (inhomogeneous) electrostriction. Conventional electrostriction yields only a small negative length change of the sample, whereas gradient electrostriction---arising from the coupling between strain and polarization gradients---produces a positive contribution nearly an order of magnitude larger and localized at the wall core. Our results demonstrate that gradient electrostriction dominates the electromechanical response of $180^{\circ}$ walls in PbTiO$_3$, supporting its inclusion in LGD models that stabilize Bloch-type domain wall structures.
	\end{abstract}
	
	\maketitle
	
	\section{Introduction}\label{sec:intro}
	
	The tensorial properties of domain walls (DWs) in ferroic materials have attracted growing interest owing to experimental and computational advances that enable the fabrication and characterization of submicron and nanoscale structures. A broad spectrum of theoretical approaches is available to describe DWs,\cite{meier2020} including first-principles calculations,\cite{Iniguez2020} machine-learned force fields,\cite{Troester2022} phase-field modeling,\cite{Voelker2011} and phenomenological Landau--Ginzburg--Devonshire (LGD) theory.\cite{Marton2010} These methods are closely connected to symmetry-based analyses using layer groups, which provide a rigorous framework for classifying DW structures and their associated tensor properties.\cite{Schranz2019,Schranz2020_JAP,Schranz2020_STO,Rychetsky2021_PZO,Schranz2022}
	
	Local polarization confined to DWs has been theoretically predicted in several perovskite systems.\cite{Stepkova2012,Marton2013} Such DW-localized order can arise from couplings including flexoelectricity,\cite{Eliseev2013} rotopolar interactions,\cite{Stengel2017,Schranz2020_STO} and biquadratic couplings between primary and secondary order parameters.\cite{Tagantsev2001} These mechanisms highlight the intricate interplay between structural distortions, polarization gradients, and strain at ferroic boundaries.
	
	In PbTiO$_3$ (PTO), the possible existence and structure of a polar ${180}^{\circ}$ DW have been extensively discussed, but the microscopic nature of the wall remains under debate. Several \emph{ab initio} studies have reported a stable Bloch-like configuration with switchable local polarization and reduced macroscopic symmetry compared to the nonpolar Ising-type wall.\cite{Meyer2002,Behera2011,Wojdel2014,Wang2017,Iniguez2020} This suggests that the transformation between Ising and Bloch DWs should be describable within the LGD framework. However, within conventional LGD models that include only standard homogeneous electrostriction, the Bloch configuration was found to be unstable.\cite{Behera2011,Wang2017,Rychet2023} To resolve this discrepancy, the LGD free energy was extended to incorporate a coupling between strain and polarization gradients of the form $e (\partial P/\partial x)^2$, referred to as gradient\cite{Hlinka_2003} or inhomogeneous electrostriction.\cite{Rychet2023} While the usual electrostriction term dominates inside the domains, the gradient coupling becomes significant within the DW and enables stabilization of the Bloch-like configuration in PTO.\cite{Rychet2023}
	
	In this work, we employ \emph{ab initio} calculations to analyze the electromechanical properties of the ${180}^{\circ}$ DW in PbTiO$_3$ and to quantify the role of inhomogeneous electrostriction. We construct a periodic equidistant array of ${180}^{\circ}$ DWs, extract layer-resolved polarization profiles and DW-induced lattice distortions, and map the results onto a continuum LGD description of electrostriction. This allows us to (i) estimate the effective tensor combination describing gradient electrostriction, (ii) separate the homogeneous and inhomogeneous electrostriction contributions to the DW-induced strain, and (iii) demonstrate that gradient electrostriction provides the dominant contribution to the longitudinal elongation localized at the DW core.
	
	\section{Array of $\mathbf{180}^{\boldsymbol{\circ}}$ domain walls}\label{sec:array}
	
	PbTiO$_3$ undergoes a uniaxial ferroelectric phase transition from the cubic to the tetragonal structure without multiplication of the unit cell. The symmetry reduction from $Pm\bar{3}m$ to $P4mm$ yields six tetragonal domain states (DSs): 
	$1_{1}\!\equiv\!(-P_s,0,0)$, $2_{1}\!\equiv\!(0,-P_s,0)$, $3_{1}\!\equiv\!(0,0,-P_s)$, and their counterparts $1_{2}$, $2_{2}$, $3_{2}$ with opposite polarization.
	
	We focus on the ${180}^{\circ}$ DW $(3_1 \,|\,\mathbf{n,p}\,|\, 3_2)$ separating the DSs $3_1\!\equiv\!(0,0,-P_s)$ and $3_2\!\equiv\!(0,0,P_s)$, with wall normal $\mathbf{n}=[1,0,0]$. The microscopic position $\mathbf{p}$ within the unit cell~\cite{Schranz2020_STO,Rychetsky2021_PZO} is energetically favored at the Pb site, at least at low temperature. In the LGD description such microscopic detail is not resolved, and the layer-group symmetry of the DW twin $(3_1 \,|\, \mathbf{n} \,|\, 3_2)$ is the four-element group
	\begin{equation}
		T_{12} = \mathbf{T}\{ 1, m_y, 2_y, \bar{1} \},
	\end{equation}
	where $\mathbf{T}$ denotes translations parallel to the DW plane.\cite{Rychet2023,Rychetsky2021_PZO} This symmetry enforces an antisymmetric N\'eel component, $P_1(x) = -P_1(-x)$, which may remain nonzero throughout the entire temperature range below $T_c$. In contrast, the Bloch component is forbidden by symmetry, since the operation $m_y$ yields $P_2(x) = -P_2(x) = 0$. A nonzero Bloch component can therefore arise only after a symmetry-lowering transition to
	\begin{equation}
		T'_{12} = \mathbf{T}\{ 1, 2_y \},
	\end{equation}
	which allows a symmetric and switchable component $P_2(x) = P_2(-x) \neq 0$.
	
	Two distinct polarization structures may thus occur in the wall: (i) an \emph{Ising-like} (\emph{Ising-N\'eel}) profile $(P_1(x), 0, P_3(x))$ with $P_2 = 0$, and (ii) a \emph{Bloch-like} profile $(P_1(x), P_2(x), P_3(x))$ with a nonzero, symmetric, and switchable Bloch component $P_2(x)$. The N\'eel component $P_1(x)$, although always present, is antisymmetric, non-switchable, and typically very small due to internal electric fields, and may often be neglected.
	
	The typical profiles are well represented by hyperbolic functions ($P_1\approx 0$):
	\begin{equation}\label{eq:profile}
		\begin{aligned}
			P_3(x) &= P_S \tanh(k x), \\
			P_2(x) &= P_B \cosh^{-1}(k x) ,
		\end{aligned}
	\end{equation}
	where $P_S$ is the spontaneous polarization, $P_B$ is the amplitude of the Bloch component, $k=2/\xi$, and $\xi$ is the DW thickness. The values of $P_S$, $P_B$, and $k$ are determined from \emph{ab initio} calculations.
	
	\begin{figure}[t]
		\centering
		\includegraphics[width=\linewidth]{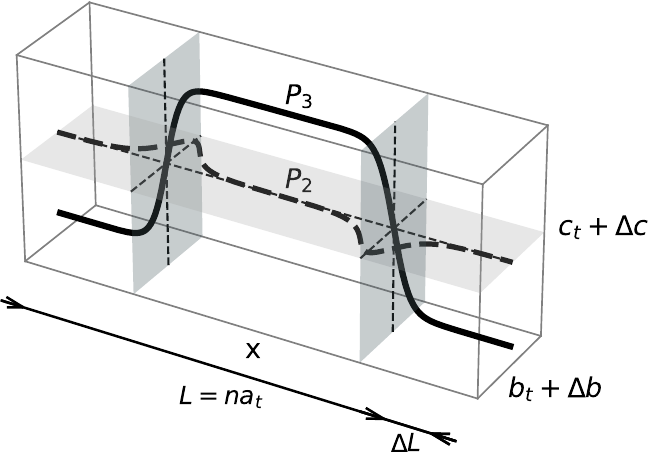}
		\caption{Schematic illustration of the periodic supercell containing two ${180}^{\circ}$ domain walls in PbTiO$_3$. The supercell is repeated along the $x$ axis, forming an equidistant array of DWs separated by $n/2$ unit cells.}
		\label{fig:struct}
	\end{figure}
	
	Because the \textsc{vasp} package (see Appendix~\ref{appendix}) enforces periodic boundary conditions, the system must be represented by a periodic model. We therefore employ a supercell containing two symmetry-related, energetically equivalent DW variants. The supercell is periodically repeated along the $x$-axis, producing an infinite, equidistant array of domain walls.
	
	The supercell comprises \( n \) unit cells arranged as \( n \times 1 \times 1 \), corresponding to a periodicity of \( n \) unit cells along the \( x \)-axis and a single unit cell along the \( y \)- and \( z \)-axes. The separation between adjacent DWs is thus \( n/2 \) unit cells.
	
	In real-space units, the tetragonal homogeneous (DW-free) reference structure has lattice dimensions \( L \times a_t \times c_t \), where \( L = n a_t \). Introducing DWs modifies the supercell dimensions to \( (L + \Delta L) \times b \times c \), with \( b = a_t + \Delta b \) and \( c = c_t + \Delta c \). The overall geometric distortion induced by the DWs can therefore be characterized by the volume change \( \Delta L \times \Delta b \times \Delta c \).
	
	To conveniently parameterize different supercell sizes and DW spacings, we define the dimensionless DW density
	\[
	\rho = \frac{2}{n},
	\]
	which gives the number of DWs per unit cell. The change of supercell dimensions generally depends on $\rho$. In the dilute limit \( n \rightarrow \infty \) (i.e., \( \rho \rightarrow 0 \)), the periodic DW interaction vanishes, corresponding to two infinitely separated and independent DWs in an effectively infinite supercell. In this regime, only the longitudinal lattice parameter is perturbed, and the dimension modification reduces to \( \Delta L \times 0 \times 0 \). Equation~(\ref{eq:profile}) then provides a good approximation to the DW profile.
	
	\section{Electromechanical properties}\label{sec:electromech}
	
	In order to describe change of the size of the supercell in Fig.~\ref{fig:struct} caused by the DWs, we use the continuum constitutive relation between strain and polarization:
	\begin{align}\label{eq: elstrict_gen}
		e_{ij} &= f_{ijkl}\frac{\partial P_k}{\partial x_l}
		+ Q_{ijkl}P_k P_l \nonumber\\
		&\quad
		+ R_{ijklmn}\frac{\partial P_k}{\partial x_m}
		\frac{\partial P_l}{\partial x_n}
		+ S_{ijkl}\sigma_{kl} .
	\end{align}
	The piezoelectricity is zero in the cubic system. Using 2-suffix notation, the flexoelectric term with 3 independent components $f_{11}$, $f_{12}$, $f_{44}$ plays minor role due to depolarizing field. The second term is the ordinary electrostriction with 3 independent components, $Q_{11}$, $Q_{12}$, $Q_{44}$. The third one is the quadratic term representing ``gradient (inhomogeneous) electrostriction'' under discussion. The tensor $R$ has 15 independent components, but considering the DW along the $x$-axis there are only 7 components contributing and $R_{ijkl11}$ with symmetric $(i, j)$ and $(k, l)$ pairs can be expressed in the 2-suffix notation:
	 \begin{equation}
		R_{ab1} =
		\begin{pmatrix}
			R_{111} & R_{121} & R_{121} & 0 & 0 & 0\\
			R_{211} & R_{221} & R_{231} & 0 & 0 & 0\\
			R_{211} & R_{231} & R_{221} & 0 & 0 & 0\\
			0 & 0 & 0 & R_{441} & 0 & 0\\
			0 & 0 & 0 & 0 & R_{551} & 0\\
			0 & 0 & 0 & 0 & 0 & R_{551}
		\end{pmatrix},
	\end{equation}
	where $a$ refers to the strain component, $b$ to the quadratic (bilinear) coupling in $P$, and the last index ``1'' to the bilinear derivative along $x$.
	
	\begin{widetext}
		For a cubic crystal with a one–dimensional polarization profile varying only
		along $\mathbf{n} = (1,0,0)$, so that all gradients reduce to
		$\partial_x$, the symmetry-allowed strain components are:
		\begin{subequations}\label{eq: elstrict_cub}
			\begin{align}
				e_1 &= f_{11}\partial_x P_1
				+ P_1^2 Q_{11} + (P_2^2 + P_3^2) Q_{12} \nonumber\\
				&\quad
				+ R_{111}(\partial_x P_1)^2 
				+ R_{121}\big[(\partial_x P_2)^2 + (\partial_x P_3)^2\big]
				+ \sigma_1 S_{11} + (\sigma_2 + \sigma_3)S_{12} ,\\[4pt]
				e_2 &= f_{12}\partial_x P_1
				+ P_2^2 Q_{11} + (P_1^2 + P_3^2) Q_{12} \nonumber\\
				&\quad
				+ R_{221}(\partial_x P_2)^2
				+ R_{211}(\partial_x P_1)^2
				+ R_{231}(\partial_x P_3)^2 
				+ \sigma_2 S_{11} + (\sigma_1 + \sigma_3)S_{12} ,\\[4pt]
				e_3 &= f_{12}\partial_x P_1
				+ P_3^2 Q_{11} + (P_1^2 + P_2^2) Q_{12} \nonumber\\
				&\quad
				+ R_{221}(\partial_x P_3)^2
				+ R_{211}(\partial_x P_1)^2
				+ R_{231}(\partial_x P_2)^2 
				+ \sigma_3 S_{11} + (\sigma_1 + \sigma_2)S_{12} ,\\[4pt]
				e_4 &= Q_{44} P_2 P_3
				+ R_{441}(\partial_x P_2)(\partial_x P_3)
				+ \sigma_4 S_{44} ,\\[4pt]
				e_5 &= f_{44}\partial_x P_3 + Q_{44} P_1 P_3
				+ R_{551}(\partial_x P_1)(\partial_x P_3)
				+ \sigma_5 S_{44} ,\\[4pt]
				e_6 &= f_{44}\partial_x P_2 + Q_{44} P_1 P_2
				+ R_{551}(\partial_x P_1)(\partial_x P_2)
				+ \sigma_6 S_{44} .
			\end{align}
		\end{subequations}
	\end{widetext}
	
	For a single $180^{\circ}$ domain wall $(3_{1}|\mathbf{n}|3_{2})$ with normal 
	$\mathbf{n}=(1,0,0)$, mechanical equilibrium requires 
	$\sigma_{1}=\sigma_{5}=\sigma_{6}=0$. \\ The strains $e_{2}$, $e_{3}$, and $e_{4}$ 
	retain their bulk values, and we neglect $P_{1}$ and the shear strains 
	$e_{5}$, $e_{6}$, consistent with the DFT results presented in the next section. 
	\begin{widetext}
	Eliminating $\sigma_{2}$ and $\sigma_{3}$ in favor of $e_{2}$ and $e_{3}$, 
	Eqs.~(\ref{eq: elstrict_cub}) reduce to
	\begin{align}\label{eq: elstrict_dw}
		e_1 &= (P_2^2 + P_3^2)Q_{12}
		+ \frac{S_{12}}{S_{11}+S_{12}}
		\Big[e_2 + e_3 - (P_2^2 + P_3^2)(Q_{11}+Q_{12})\Big]  \nonumber\\
		&\quad
		+ \Big[(\partial_x P_2)^2 + (\partial_x P_3)^2\Big]
		\Big(R_{121} - \frac{S_{12}}{S_{11}+S_{12}}(R_{221}+R_{231})\Big)
		+ (\sigma_2 + \sigma_3)S_{12}, \\
		e_2 &= e_{2,S} = P_S^2 Q_{12},\qquad  
		e_3 = e_{3,S} = P_S^2 Q_{11},\qquad  
		e_4 = e_5 = e_6 = 0 .
	\end{align}
\end{widetext}
	Only the longitudinal strain $e_{1}(x)$ varies across the wall.  
	In the bulk domain $e_{1,S}=P_S^2 Q_{12}$, and the wall-induced strain deviation  
	$\Delta e_{1} \equiv e_1 - e_{1,S}$ separates as
	\begin{align}\label{eq:delta_e1}
		\Delta e_1 &= \Delta e_{1,eh} + \Delta e_{1,en} \nonumber\\
		&= \big[(P_{2,S}^2-P_2^2)+(P_{3,S}^2-P_3^2)\big] I_Q \nonumber\\
		&\quad
		- \big[(\partial_x P_2)^2+(\partial_x P_3)^2\big] I_R ,
	\end{align}
	with
	\begin{align}\label{eq:IQR}
		I_Q &= \frac{S_{12}}{S_{11}+S_{12}}\,(Q_{22}+Q_{23}) - Q_{12},\\
		I_R &= \frac{S_{12}}{S_{11}+S_{12}}\,(R_{221}+R_{231}) - R_{121}.
	\end{align}
	The corresponding change in sample length is
	\begin{equation}
		\Delta \mathcal{L}
		= \int_{-\infty}^{+\infty} \Delta e_1(x)\,dx
		= \Delta \mathcal{L}_{eh} + \Delta \mathcal{L}_{en}.
	\end{equation}
	For a supercell containing two well-separated walls described by  
	Eq.~(\ref{eq:profile}), the total length change is
	\begin{align}\label{eq:deltaL}
		\Delta L &= \Delta L_{eh}+\Delta L_{en}
		= 2\Delta \mathcal{L} \nonumber\\
		&= \frac{4}{k}(P_S^2 - P_B^2)\,I_Q 
		-\frac{8k}{3}\!\left(P_S^2+\frac{P_B^2}{2}\right)I_R ,
	\end{align}
	where $k=2/\xi$.  
	Equation~(\ref{eq:deltaL}) applies in the long-supercell limit, where the two
	walls do not interact ($\rho \approx 0$) and the transverse lattice parameters
	also remain unchanged, $\Delta b = \Delta c = 0$ (see Fig.~\ref{fig:struct}). In Eqs.~(\ref{eq:delta_e1}) and (\ref{eq:IQR}) the conventional and gradient electrostriction terms contain $I_Q$ and $I_R$, respectively. 
	
	\section{Ab initio model}\label{sec:abinitio}
	
	Equation~(\ref{eq:deltaL}) expresses the domain–wall–induced length change
	$\Delta L$ in terms of the spontaneous polarization $P_S$, the Bloch
	amplitude $P_B$, the inverse wall thickness $k$, and the effective
	combinations $I_Q$ and $I_R$ of electrostrictive and gradient–electrostrictive
	coefficients.  The purpose of this section is to show how these quantities are
	obtained from first-principles calculations for PbTiO$_3$ and how they connect
	to the continuum description of Sec.~\ref{sec:electromech} (\textit{Electromechanical properties});
	technical details of the electronic-structure calculations are given in the
	Appendix.
	
	First, the bulk elastic compliances $S_{11}$ and $S_{12}$ (listed as
	$S_{11}$ and $S_{12}$ in Table~\ref{tab:table}) and the electrostriction coefficients
	$Q_{11}$ and $Q_{12}$ were determined for cubic PbTiO$_3$ from homogeneous
	strain states.  The spontaneous polarization $P_S$ was obtained from the fully
	relaxed tetragonal ferroelectric phase.  These bulk calculations provide the
	input for the homogeneous electrostriction contribution to
	Eq.~(\ref{eq:deltaL}) and allow us to evaluate $I_Q$ via
	Eq.~(\ref{eq:IQR}).  The resulting values of $S_{ij}$, $Q_{ij}$, $I_Q$, and
	$P_S$ are summarized in Table~\ref{tab:table}.
	As the reference structures we further consider homogeneous tetragonal supercells $n\times 1\times 1$ with $n = 8,10,\ldots,22$, and get relaxed lattice parameters $L\times a_t\times c_t$, $L=n a_t$. 
	
\begin{figure}[t]
		\centering
		\includegraphics[width=\linewidth]{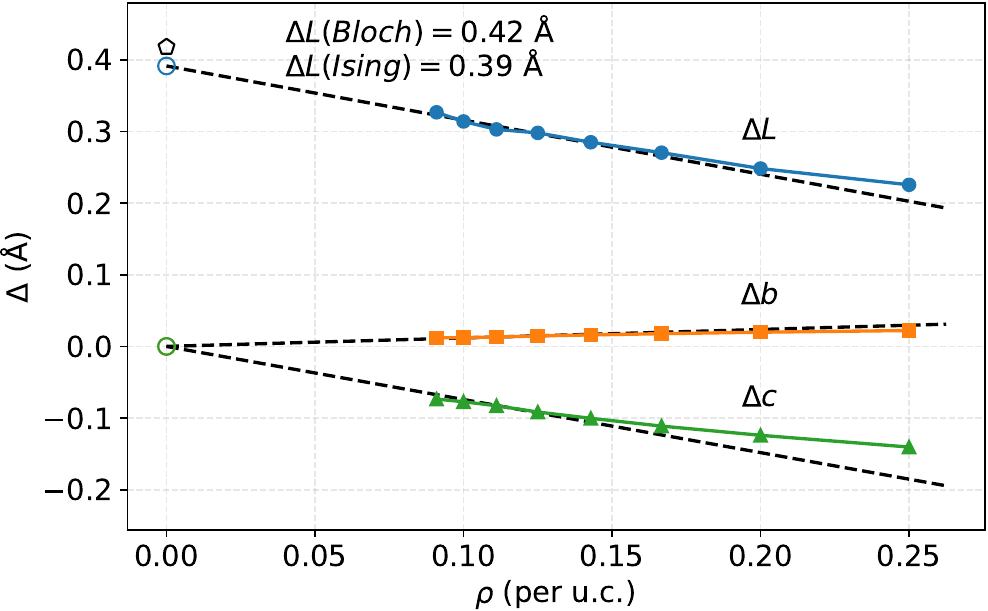}
		\caption{(Color online) The longitudinal elongation $\Delta L$, as well as the transverse expansion $\Delta b$ and contraction $\Delta c$, are shown as functions of $\rho$ for the Ising domain wall (DW). The corresponding Bloch DW curves are very similar; only $\Delta L$ for the Bloch case at $\rho = 0$ is shown. The values for independent DWs occur at $\rho = 0$.}
		\label{fig:cell_sizes_diff}
	\end{figure}

	To access the domain–wall contribution, we then considered the periodic array
	of $180^\circ$ DWs introduced in Sec.~\ref{sec:array} (\textit{Array of
		$\mathbf{180}^{\boldsymbol{\circ}}$ domain walls}).  Each supercell of size
	$n\times 1\times 1$ contains two symmetry-related DWs separated by $n/2$ unit cells along
	the $x$ axis, corresponding to a wall density
	$\rho = 2/n$.  For each $n$, a full structural relaxation under zero external
	stress was performed, yielding the relaxed $\rho$-dependent lattice parameters of the supercell 
	$L+\Delta L(\rho)\times a_t+\Delta b(\rho)\times c_t+\Delta c(\rho)$ and the microscopic ionic displacements.  
	Comparing it with the homogeneous tetragonal reference supercell
	containing the same number of unit cells, the supercell size change $\Delta L(\rho)$, $\Delta b(\rho)$ and $\Delta c(\rho)$ was calculated, see Figs.~\ref{fig:struct} and \ref{fig:cell_sizes_diff}. 
	
\begin{figure}[t]
		\centering
		\includegraphics[width=\linewidth]{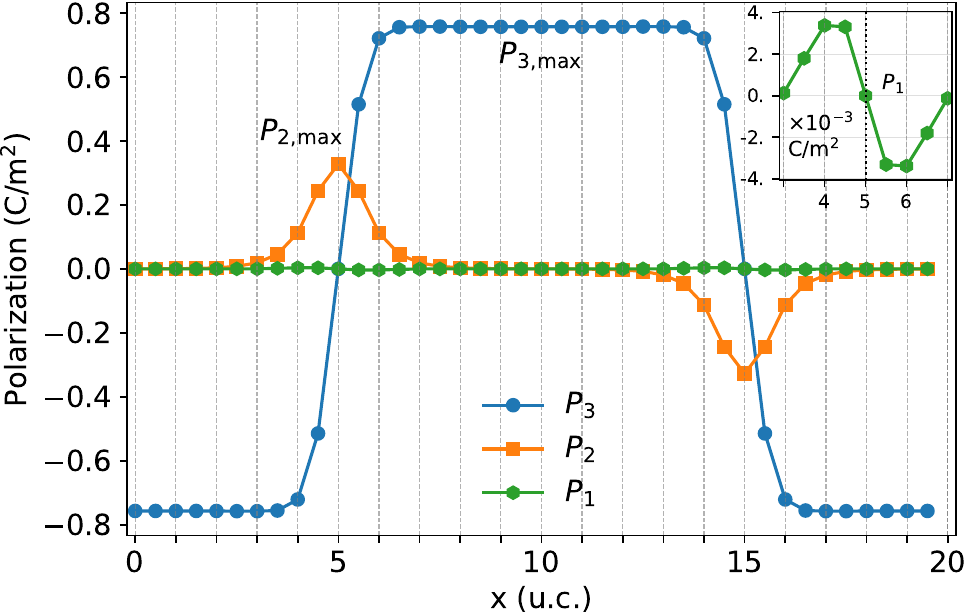}
		\caption{(Color online) Layer-resolved polarization components across the Bloch-type ${180}^{\circ}$ DW for $n=20$. Integer indices correspond to Pb planes, while half-integer indices (e.g., 0.5, 1.5, \dots) correspond to Ti planes.}
		\label{fig:profile}
	\end{figure}

\begin{figure}[t]
		\centering
		\begin{overpic}[width=\columnwidth]{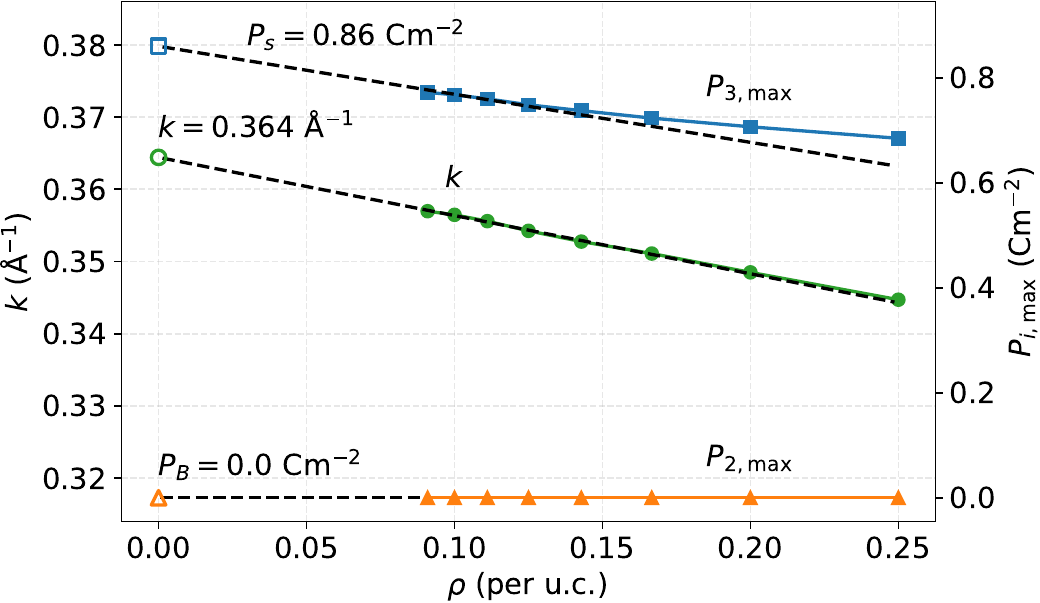}
			\put(80,52){\small\textbf{(a)}}
		\end{overpic}
		
		\vspace{0.6em}
		
		\begin{overpic}[width=\columnwidth]{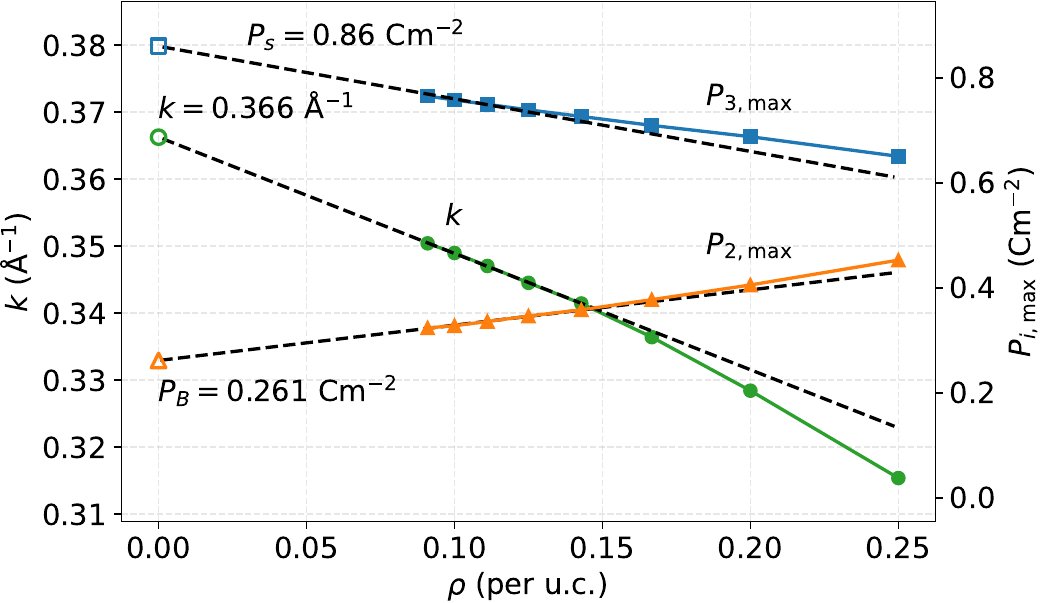}
			\put(80,52){\small\textbf{(b)}}
		\end{overpic}
		\caption{(Color online) Dependence of the domain polarization $P_{3,\mathrm{max}}$, the Bloch component $P_{2,\mathrm{max}}$, and the inverse DW thickness $k$ on the DW density $\rho$. \textbf{(a)} Ising case; \textbf{(b)} Bloch case. 
			The values for independent DWs occur at $\rho = 0$.}
		\label{fig:slopes}
	\end{figure}

	From the relaxed structures we extracted the layer-resolved polarization
	components $P_1(x)$, $P_2(x)$, and $P_3(x)$ across the supercell. An example of
	the relaxed Bloch profile for $n=20$ is shown in Fig.~\ref{fig:profile}; the
	Ising profile is similar, with $P_2=0$; compare also Refs.~\cite{Rychet2023,Chege2025}.
	The points represent the positions of alternating Pb and Ti planes, the DW centers are at the Pb-planes corresponding to the lowest energy. 
	These profiles 
	provide, for each wall density $\rho$, the maximum polarization 
	$P_{3,\max}(\rho)$ in the domain, the amplitude $P_{2,\max}(\rho)$ of the Bloch component, and the inverse wall thickness $k(\rho)$, defined as $k(\rho)=P_{3,\max}^{-1}\partial_xP_3(x)\rvert_{x=0}$.
	So we have 6 quantities characterizing DWs array: 
	the profile parameters 
	$P_{3,\max}(\rho)$, $P_{2,\max}(\rho)$ and $k(\rho)$ shown in Fig.~\ref{fig:slopes} and 
	the supercell size changes $\Delta L(\rho)$, $\Delta b(\rho)$ and $\Delta c(\rho)$ plotted in Fig.~\ref{fig:cell_sizes_diff}. To get description of isolated DWs we need to extrapolate the curves to $\rho=0$ yielding parameters in Eq.~(\ref{eq:deltaL}):  $P_{3,\max}(\rho)\to P_S$, $P_{2,\max}(\rho)\to P_B$ and $k(\rho)\to k$, $\Delta L(\rho)\to \Delta L$, $\Delta b(\rho)\to 0$ and $\Delta c(\rho)\to 0$,  Table~\ref{tab:table}. The extrapolations where done by linear fits of the lowest 4 points. The only unknown quantity, the gradient electrostriction coefficient $I_R$, is calculated from Eq.~(\ref{eq:deltaL}). 
	
	\newcommand{\E}[1]{\times10^{#1}}
	
	\begin{table}[t]
		\caption{Material constants (independent of the DWs) and the characteristics of the Ising and Bloch domain walls. Most quantities were obtained from \emph{ab initio} calculations, while $\Delta L_{eh}$, $\Delta L_{en}$, and $I_R$ were derived from the continuum model, Eq.~(\ref{eq:deltaL}).}
		\label{tab:table}
		\begin{ruledtabular}
			\begin{tabular}{l c l}
				Component & Value & Unit \\
				\hline
				\multicolumn{3}{c}{\textbf{Material constants}} \\
				$S_{11}$ & $3.998\E{-3}$ & GPa$^{-1}$ \\
				$S_{12}$ & $-1.128\E{-3}$ & GPa$^{-1}$ \\
				$S_{66}$ & $9.845\E{-3}$ & GPa$^{-1}$ \\
				$Q_{12}$ & $-1.857\E{-2}$ & m$^{4}$C$^{-2}$ \\
				$Q_{11}$ & $7.900\E{-2}$ & m$^{4}$C$^{-2}$ \\
				$I_Q$    & $-5.181\E{-3}$ & m$^{4}$C$^{-2}$ \\
				$I_R$    & $-6.019\E{-21}$ & m$^{6}$C$^{-2}$ \\
				$P_S$    & $0.86$          & C\,m$^{-2}$ \\
				\hline
				\noalign{\vskip 2pt}
				\multicolumn{3}{c}{\textbf{Ising DW}} \\
				$\Delta L$        & $3.9\E{-11}$  & m ($0.39\,\text{\AA}$) \\
				$\Delta L_{eh}$   & $-0.42\E{-11}$ & m ($-0.042\,\text{\AA}$) \\
				$\Delta L_{en}$   & $4.3\E{-11}$   & m ($0.43\,\text{\AA}$) \\
				$k$               & $3.64\E{9}$    & m$^{-1}$ ($0.364\,\text{\AA}^{-1}$) \\
				\hline
				\noalign{\vskip 2pt}
				\multicolumn{3}{c}{\textbf{Bloch DW}} \\
				$\Delta L$        & $4.2\E{-11}$  & m ($0.42\,\text{\AA}$) \\
				$\Delta L_{eh}$   & $-0.38\E{-11}$ & m ($-0.038\,\text{\AA}$) \\
				$\Delta L_{en}$   & $4.6\E{-11}$   & m ($0.46\,\text{\AA}$) \\
				$k$               & $3.66\E{9}$    & m$^{-1}$ ($0.366\,\text{\AA}^{-1}$) \\
				$P_B$             & $0.26$         & C\,m$^{-2}$ \\
			\end{tabular}
		\end{ruledtabular}
	\end{table}
	
	\section{Discussion}
	
	The dependencies of various quantities on the DW density in the multidomain sample, together with their fits, allow us to extract very small yet reliable values, such as a longitudinal elongation on the order of a fraction of an ångström. At finite DW density $\rho$, the walls exert forces (averaged over the supercell) on the bulk lattice, expanding it along the $b$-axis and compressing it along the polar $c$-axis. This reduces the tetragonality and leads to a decrease in the domain polarization, $(P_{3,\max} < P_S)$. As $\rho \to 0$, the transverse dimensions remain unchanged $(\Delta b = \Delta c = 0)$, and the longitudinal elongation $\Delta L$, originating from two independent DWs, remains only a fraction of an ångström.
	
	The continuum electromechanical equation, Eq.~(\ref{eq: elstrict_gen}), enables us to separate the contributions of conventional and gradient electrostriction to $\Delta L$, as expressed in Eq.~(\ref{eq:deltaL}). Electrostriction in the standard LGD framework is governed solely by the tensor $Q$, and in our case its contribution $\Delta L_{eh}$ is very small and negative; thus, it cannot account for the observed positive elongation $(\Delta L > 0)$. The remaining, dominant contribution arises from gradient electrostriction, $\Delta L_{en}$, which is positive and an order of magnitude larger than the small negative $\Delta L_{eh}$ (see Table~\ref{tab:table}). This underscores the crucial role of gradient electrostriction at the DW center. Consistently, the electrostriction parameter $I_R$ extracted from Eq.~(\ref{eq:deltaL}) is identical for both Ising and Bloch DWs.
	
	The difference in elongation between Ising and Bloch DWs is very small, $\approx 0.03\ \text{\AA}$, indicating a weak dependence on the precise DW profile. Accordingly, the extremely small N\'eel component $P_1 \approx 10^{-3}\,\mathrm{C/m^2}$ has negligible influence (see inset in Fig.~\ref{fig:profile}). The longitudinal elongation $\Delta L$ is highly localized: about $97\%$ of the effect originates within two unit cells at the DW center.
	
	\section{Summary}
	
	We have investigated the electromechanical properties of the ${180}^{\circ}$ ferroelectric domain wall in PbTiO$_3$ by combining first-principles calculations with a continuum Landau--Ginzburg--Devonshire description that includes inhomogeneous electrostriction. Symmetry analysis using layer groups shows that, in the high-symmetry DW state, only an antisymmetric N\'eel component of the polarization is allowed, while the symmetric and switchable Bloch component requires a symmetry-lowering transition within the DW.
	
	Using density-functional theory, we constructed periodic arrays of ${180}^{\circ}$ DWs in PbTiO$_3$ with varying wall density, relaxed the atomic and lattice degrees of freedom, and extracted layer-resolved polarization profiles and DW-induced lattice distortions. By analyzing the dependence on DW density in the dilute limit, we obtained the effective spontaneous polarization, Bloch amplitude, wall thickness, and longitudinal elongation associated with a single isolated DW, for both Ising-like and Bloch-like configurations.
	
	Mapping these results onto the continuum constitutive relation~(\ref{eq: elstrict_gen}) allowed us to separate the contributions of conventional (homogeneous) electrostriction and gradient electrostriction to the DW-induced length change. We find that the conventional electrostriction contribution is small and negative, whereas the gradient-electrostriction contribution is positive and nearly an order of magnitude larger, leading to a net elongation that is strongly localized within two unit cells around the DW core. The extracted gradient-electrostriction parameter,  $I_R\approx 10^{-21} \mathrm{C^{-2}m^6}$, is nearly identical for Ising and Bloch DWs, and the DW-induced strain is only weakly sensitive to the precise profile shape or the tiny N\'eel component. This value also provides a typical magnitude for the components of the tensor $R$, consistent with the value adopted in the phenomenological model of Ref.~\cite{Rychet2023}. 
	
	These results provide direct quantitative support for the central role of gradient (inhomogeneous) electrostriction in the electromechanical response \cite{Hlinka_2003} and structural stabilization of ${180}^{\circ}$ domain walls in PbTiO$_3$ \cite{Rychet2023}. More broadly, they demonstrate how combining \emph{ab initio} calculations with continuum modeling enables a systematic determination of gradient couplings in ferroic DWs.

	\section*{Acknowledgments}
	
	This work was supported by the Czech Science Foundation (GAČR), project No.\ 25-18870L. 
	W.S.\ gratefully acknowledges support from the Austrian Science Fund (FWF), project No.\ 10.55776/PIN2246224. 
	Computations were performed using MetaCentrum resources supported by the e-INFRA CZ project (ID: 90254).
	
	\appendix
	
	\section{Computational details}\label{appendix}
	First-principles calculations were performed using Kohn–Sham density-functional theory as implemented in the VASP package 
	 \cite{Kresse1996CMS,Kresse1996PRB,Kresse1999PAW}. The interaction between
	valence electrons
	and ionic cores was treated within the projector-augmented-wave (PAW)
	method, while
	exchange–correlation effects were described using the PBEsol generalized-gradient approximation.
	The PAW datasets treated 12 valence electrons for Ti
	($3s^{2}\,3p^{6}\,3d^{2}\,4s^{2}$),
	14 for Pb ($5d^{10}\,6s^{2}\,6p^{2}$), and 6 for O ($2s^{2}\,2p^{4}$).
	A plane-wave kinetic-energy cutoff of $\mathrm{ENCUT}=520$~eV and
	$\mathrm{PREC}=\mathrm{Accurate}$
	were used throughout. Brillouin-zone integrations employed a $1\times
	6\times 6$ Monkhorst--Pack
	mesh, 
	chosen to reflect the elongated supercell geometry, with the long axis along $x$.

	Structural relaxations were performed using the conjugate-gradient algorithm
	(IBRION = 2) for up to 300 ionic steps (NSW = 300).  Both the atomic positions
	and the cell shape and volume were allowed to relax (ISIF = 3) under zero
	external stress.  The electronic self-consistency criterion was set to
	EDIFF = $10^{-8}$~eV, and the ionic relaxations were converged when all
	Hellmann--Feynman forces were smaller than 0.001~eV\,Å$^{-1}$
	(EDIFFG = $-0.001$~eV\,Å$^{-1}$).  Gaussian smearing with ISMEAR = 0 and
	$\sigma = 0.02$~eV was employed, and real-space projectors were disabled
	(LREAL = \texttt{.FALSE.}) to ensure high accuracy.
	
	The $180^\circ$ domain walls were modeled using supercells containing
	$n$ Pb, $n$ Ti, and $3n$ O atoms, corresponding to $n\times 1\times 1$
	tetragonal units with $n$ between 8 and 22, and lattice vectors of the form
	$L(=n a_t) \times a_t \times c_t$ in the homogeneous reference structure.  Periodic
	boundary conditions enforce an equidistant array of two symmetry-related walls
	along the $x$ direction.  For each $n$, the relaxed lattice parameters and
	ionic positions were used to determine the supercell length change
	$\Delta L(\rho)$ and the layer-resolved polarization components across the
	wall.  Local polarization profiles were constructed from the relative
	cation--anion displacements with respect to the cubic structure with the help of Born effective charges.
	
	\bibliography{pto_paper}
	
\end{document}